# Enhancing Cluster Quality of Numerical Datasets with Domain Ontology

Sudath Rohitha Heiyanthuduwage
*School of Computing, Mathematics and Engineering*
Charles Sturt University
Australia
SHeiyanthuduwage@csu.edu.au

Md Anisur Rahman
*School of Computing, Mathematics and Engineering*
Charles Sturt University
Australia
ARahman@csu.edu.au

Md Zahidul Islam
*School of Computing, Mathematics and Engineering*
Charles Sturt University
Australia
ZIslam@csu.edu.au

*Abstract*— **Ontology-based clustering has gained attention in recent years due to the potential benefits of ontology. Current ontology-based clustering approaches have mainly been applied to reduce the dimensionality of attributes in text document clustering. Reduction in dimensionality of attributes using ontology helps to produce high quality clusters for a dataset. However, ontology-based approaches in clustering numerical datasets have not been gained enough attention. Moreover, some literature mentions that ontology-based clustering can produce either high quality or low-quality clusters from a dataset. Therefore, in this paper we present a clustering approach that is based on domain ontology to reduce the dimensionality of attributes in a numerical dataset using domain ontology and to produce high quality clusters. For every dataset, we produce three datasets using domain ontology. We then cluster these datasets using a genetic algorithm-based clustering technique called GenClust++. The clusters of each dataset are evaluated in terms of Sum of Squared-Error (SSE). We use six numerical datasets to evaluate the performance of our ontology-based approach. The experimental results of our approach indicate that cluster quality gradually improves from lower to the higher levels of a domain ontology.**

*Keywords— Clustering, Domain Ontology, Cluster Evaluation, Genetic Algorithm, Data Mining.*

## I. INTRODUCTION

Clustering refers to grouping the records in a dataset into two or more clusters so that similar records are grouped together into a single cluster and dissimilar records are grouped into different clusters [1]. Applications of data clustering varies from analysing retail data to business, weather, medical and many other data to identify specific and useful data patterns in them [2]. Generally, data clustering is done on datasets with a large number of records and a large number of attributes. Because of that clustering results also include many records with many attributes and are generally difficult to be interpreted without some domain knowledge. Ontology has been proposed as an option for both enhancing clustering results and as a means of understanding clustering results. An ontology has been defined as a formal specification of a specific domain [3]. An ontology defined for a specific domain includes concepts, their attributes, relationships among the concepts, and various constraints of that domain. Domain ontologies have been applied in different domains such as learning, medicine, engineering and travel. The domain knowledge specified in a specific ontology helps us to get a comprehensive understanding of that domain.

Current ontology-based clustering approaches have focused on different aspects of cluster analysis. They include use of ontology information for data pre-processing, enriching term vectors, re-weight the vectors, etc. [4]. In clustering process, ontology could play a role in pre-processing stage or in clustering stage or in a later stage in understanding the clustering results. For example Hotho et al. [5] has proposed a technique to use an ontology for pre-processing data using an algorithm that over performs other data pre-processing techniques. Some approaches have focused on applying ontology directly on clustering algorithms and attempted to improve the clustering process. An algorithm named Onto-SVD (Singular Value Decomposition) has been used to find the semantic similarity between stored memories in [6]. Onto-SVD uses SVD to identify topics with the help of semantic similarity and semantic feature selection with k-means.

Many ontology-based clustering approaches have focused on text document clustering. They have analysed text documents with a higher number of terms and used ontology to reduce dimensionality of attributes, for example [5, 7, 8]. An ontology-based framework that applies ontology to organize text documents considering the cohesive groups of segment-based portions has been designed in [9]. An ontology-based genetic algorithm has been applied to cluster text documents in [10]. These techniques initially represent the nouns found in a text document as a two dimensional structure or Vector Space Models (VSM) of the document and frequency of words found in each document. They reduce dimensionality of attributes by domain specific nouns instead of general terms [10].

Some literature shows that ontology-based clustering increase cluster quality while others show that ontology-based clustering does not increase cluster quality of a dataset [7]. Moreover, ontology-based clustering techniques on numerical datasets and domain ontology-based techniques have not gained enough attention. Therefore, in this paper we present a domain ontology-based clustering approach that alleviates the limitations of ontology-based clustering that are mentioned in the above paragraph. In our approach, for every dataset we produce three datasets using domain ontology. We then cluster these datasets using a parameter free genetic algorithm-based clustering technique called GenClust++ [11]. The clusters of each dataset are evaluated in terms of Sum of Squared-Error (SSE). We use six numerical datasets to evaluate the performance of our proposed ontology-based clustering technique. The experimental results of our approach indicates a gradual improvement in cluster quality from lower to the higher levels of ontology.

The rest of this paper is organised as follows. Section 2 presents the related work and section 3 introduces our approach. Section 4 elaborates on experimental results and provides a discussion on them. Section 5 concludes the paper.

## II. RELATED WORK

Data clustering is applied on different types of data including numerical, categorical and text data. Ontology-based clustering methods have predominantly been applied in clustering text documents. The problems addressed in ontology-based text document clustering have been categorised into four main categories in [12]. 1. Synonym (having several words with the same meaning) and polysemy (one word having several different meanings) problems and solving them using approaches for Word-Sense Disambiguation (WSD). 2. High dimensionality of terms increases the processing time and decrease the cluster quality. 3. Extracting core semantic features from the text that is useful to improve cluster quality. 4. Providing distinguishable and meaningful descriptions to clustering results that is useful for users to understand the results.

Both polysemous and synonymous nouns have been identified as relatively prevalent and considered to be fundamentally significant in document cluster formation in [7]. When synonyms nouns are found in a document, they are considered to be a single concept of an ontology. On the other hand, when polysemous nouns are found, they are considered to be different concepts of an ontology. Their work [7] has demonstrated that nouns identified in documents themselves help to perform better clustering. Their work has shown the importance of polysemous and synonymous nouns in clustering.

While synonymy leads to reducing the dimensionality of attributes, polysemy leads to increasing the dimensionality of attributes [7]. Typically, text clustering is considered to be difficult in practical settings as it involves clustering in a high dimensional space [5]. Hotho et al. [5], have used a domain ontology to reduce the dimensionality of attributes with the help of an algorithm based on some heuristic rules. This algorithm has helped to identify core semantic features as well. Extracting core semantic features relevant to the domain help to exclude the nouns that are irrelevant to a specific domain. Indenting core semantic features not only further reduce the dimensionality of features, but also significant in providing domain specific knowledge to the users. It has been found that higher cluster purity can be obtained using core semantic features compared to using polysemous/synonymous nouns and using all nouns [7].

Two popular ontologies that have been used in ontology-based clustering are WordNet ontology and Gene ontology (GO). WordNet ontology has mostly been used as reference ontology in document clustering approaches while GO has been used as a domain ontology specific to medical domain. A number of ontology-based clustering approaches that use WordNet have been listed in [7]. WordNet has also been used in [13] to identify the relationships between the concepts of an ontology. Then, an ontology index has been generated based on the relationships between concepts and their sub concepts. The ontology and ontology index have been deployed in ontology-based weighted clustering to overcome bottleneck issues found in Hadoop in handing big data [13].

In the medical domain clustering has been applied on gene data to identify specific genes associated to specific diseases. GO has been proposed as a means of annotating the results of microarray experiments. An R-based fast software (csbl.go) has been proposed for advanced gene annotation in [14]. A document is defined as a structured set of fields in [15] that includes references to other databanks such as Gene Ontology (GO) and text that provide information about organisms. The query results are presented to the users in hierarchical folders with a string label and the cluster size. They [15] have applied ontology-based hierarchical clustering to make searching the databanks efficient and effective.

Increase in number of web services makes a user finding the favoured web services efficiently and accurately challenging. The method proposed for ontology-based clustering in [16] attempt to overcome this problem. As a solution, an ontology based on similarity and specificity of web services is generated. The ontology is then used to avoid incorrectly placing web services in clusters and achieve a higher clustering performance.

## III. OUR PROPOSED METHOD: ENHANCING CLUSTERING QUALITY WITH DOMAIN ONTOLOGY

In this section we present our approach to ontology-based clustering that is based on domain ontology. Our methodology includes six steps: (1) Deriving a domain ontology from a dataset, (2) Normalising dataset, (3) Creating datasets from each level of domain ontology for clustering, (4) Clustering each level of dataset and (5) Evaluating the clustering results. In our approach we focus on performing data clustering on numerical datasets. Another aspect we consider is purpose of the dataset and how easy for us to understand the application domain. That would make it easy for us to generate a domain ontology for each of them. We also checked whether a dataset has enough attributes in it, neither too few nor too many. The datasets we have selected have a range of ten to thirty-three attributes. The reason for us to avoid datasets with few attributes is that would become difficult for us to get several levels in a domain ontology. If a dataset includes too many attributes, the domain ontology we create would have too many levels. That would create ontologies with different and incomparable levels that would avoid a fair comparison of cluster quality of the datasets we generate based on the levels of the ontology. In below subsections, we explain each step of our methodology.

### A. Step 1: Deriving a Domain Ontology from a Dataset

Many existing approaches have used WordNet [17] as a reference ontology. WordNet is a lexical database for English and it include many general terms that make a tall hierarchy of concepts in ontology. In our approach, we propose to create domain specific ontology for each dataset that alleviate the use of general concepts. When we create a domain ontology for a specific dataset, we consider all the attributes in the dataset as the lowest level concepts of the domain ontology. Then, we see whether these lowest level concepts belong to a common type or kind that would form a higher level concept. That would create an inclusion relationship [18] between several lowest level concepts and a higher level concept. Inclusion relationship can be considered as a parent-child relationship or a type-of or a kind-of relationship. This process we follow to identify higher level (parent) concepts is a heuristic approach. In below subsections we elaborate on creating a domain ontology for Travel Review (TR) dataset to provide details of this process we follow to create a domain ontology. We then apply the same process to the other datasets as well and create domain specific ontologies for all the six datasets.

TABLE I. INITIAL SET OF ATTRIBUTES AND LEVEL 1 CONCEPTS OF THE ONTOLOGY

| # | Attribute | # | Attribute | # | Attribute |
|---|---|---|---|---|---|
| 1 | Churches | 9 | restaurants | 17 | swimming pools |
| 2 | Resorts | 10 | pubs/bars | 18 | gyms |
| 3 | Beaches | 11 | local services | 19 | bakeries |
| 4 | Parks | 12 | burger/ pizza shops | 20 | beauty & spas |
| 5 | Theatres | 13 | Hotels & other lodgings | 21 | cafes |
| 6 | Museums | 14 | juice bars | 22 | viewpoints |
| 7 | Mall | 15 | art galleries | 23 | monuments |
| 8 | Zoos | 16 | dance clubs | 24 | gardens |

*1) Identifying Level 1 Concepts of a Domain Ontology*

This is the first step in our process we follow to create a domain ontology. All the attributes except sequential numbers, attributes with the same value for all the records and class attributes (if exists) are considered to be in level 1 of the ontology. For example, TR dataset includes 25 attributes, yet the first attribute, user id is a sequential number. Therefore, except that all the other attributes are listed in Table I. These 24 attributes form the level 1 of the domain ontology named *Travel Review Ontology (TRO)* that specifies the domain knowledge of TR domain. These 24 attributes in Table I are used in the next step to form level 2, the next level of TRO.

*2) Deriving the Level 2 Concepts of the Ontology*

Here we see what lowest level concepts belong to what common higher level concepts and take those higher level concepts as the level 2 concepts of TRO. Higher level concepts are defined based on ontology engineers understanding of the domain. We consider that the concepts *Churches, Museums, Art Galleries* and *Monuments* belong to the higher level concept *Cultural Place* (Table II). The concepts *Beaches* and *Viewpoints* belong to the higher level concept *Natural Place*; the concepts *Parks, Malls, Zoo* and *Gardens* belong to the higher level concept *Natural Place*. The concepts *Parks, Malls, Zoo* and *Gardens* belongs to the higher level concept *Social Place*.

TABLE II. LEVEL 2 CONCEPTS IDENTIFIED FROM LEVEL 1 CONCEPTS

| # | Level 2 concepts | Level 1 concepts |
|---|---|---|
| 1 | Cultural Places | Churches, Museums, Art Galleries, Monuments |
| 2 | Natural Place | Beaches, Viewpoints |
| 3 | Social Place | Parks, Malls, Zoo, Gardens |
| 4 | Accommodation | Resorts, Hotels and Other Lodgings |
| 5 | Entertainment | Theatres, Dance Clubs |
| 6 | Food and Drinks | Restaurants, Pubs and Bars, Burger and Pizza Shops, Juice Bar, Bakeries, Cafes |
| 7 | Services | Local Services, Swimming Pools, Gyms, Beauty and Spa |

The higher level concept *Accommodation* includes the lower level concepts *Resorts, Hotels and Other Lodgings*. The higher level concept *Entertainment* includes the lover level concepts *Theatres* and *Dance Clubs*. The higher level concept *Food and Drinks* includes the lower level concepts *Restaurants, Pubs and Bars, Burger* and *Pizza Shops, Juice Bar, Bakeries* and *Cafes*. The higher level concept *Services* includes the lower level concepts *Local Services, Swimming Pools, Gyms, Beauty and Spa* as shown in Table II. In next step, we use these seven level 2 concepts to identify the level 3 concepts of TRO.

*3) Deriving the Level 3 Concept of the Ontology*

Again, the seven concepts identified for level 2 of TRO are used to form two level 3, higher level concepts, *Attraction* and *Facility* in TRO. The level 2 concepts, *Cultural Place, Natural Place* and *Social Place* belong to the higher level concept *Attractions*. The level 2 concepts, *Accommodation, Entertainment, Food and Drinks* and *Services* belong to the higher level concept *Facility* as shown in Table III. Now the level 3 of TRO has only 2 concepts and we are unable to go any further higher to form another level of TRO. These two level 3 concepts are now considered to be the child concepts of the root concept of the ontology, *Thing*. Even though we get only 3 levels of concept hierarchy in domain ontology for these datasets, an ontology engineer may get different number of levels in forming domain ontologies for another dataset.

TABLE III. LEVEL 3 CONCEPTS IDENTIFIED FROM LEVEL 2 CONCEPTS

| # | Level 3 Concepts | Level 2 Concepts |
|---|---|---|
| 1 | Attraction | Cultural Place, Natural Place, Social Place |
| 2 | Facility | Accommodation, Entertainment, Food and Drinks, Services |

*4) Travel Review Ontology*

When these inclusion relationships among the concepts at all the three levels are considered, we get a domain ontology for *TR* dataset. TRO ontology we created in the ontology editing tool Protégé [19] is given in Fig. 1. In that *owl:Thing* is the root concept and the prefix *owl:* in root concept refers to Web Ontology Language (OWL) [20].

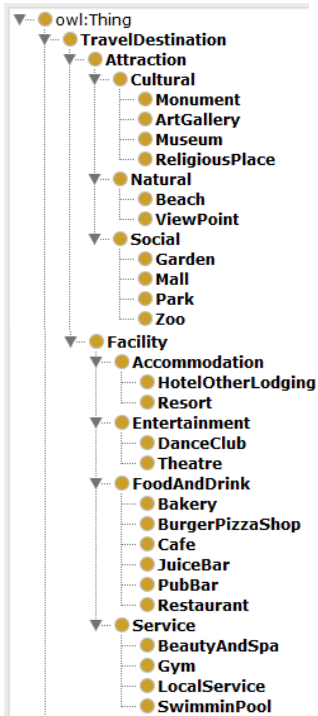

Fig. 1. Domain ontology for *TR* dataset.

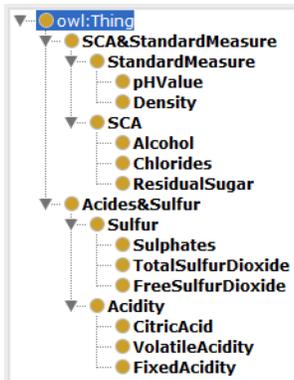

Fig. 2. Domain ontology for *WQ* dataset.

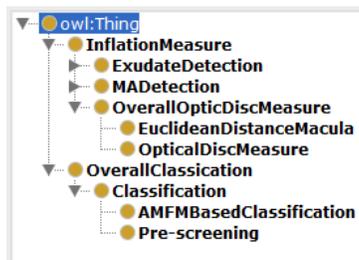

Fig. 3. Domain ontology for *DRD* dataset.

*1) Domain Ontologies for Each Dataset*

We follow the same approach we followed and repeat the steps 3.1.1 to 3.1.5 to derive domain ontologies for other datasets as well. For example, Fig. 2 and Fig. 3 show the domain ontologies developed for Wine Quality (WQ) and Diabetic Retinopathy Debrecen (DRD) datasets. Each of the 6 ontologies includes three levels of inclusion hierarchy. That makes all the six domain ontologies comparable to each other specially when they are used for clustering the datasets.

*B. Step 2: Normalising the Datasets*

Normalising data is required on some datasets to simplify the data to make them suitable for clustering and the clustering results comparable. All the original datasets we use in this work are numerical. Still, each attribute of them had different scope of values, some values were below 0 and some values were above 100. Therefore, attribute values in each dataset were normalized to make all values between zero and one. A simple way of normalising a value of an attribute is to divide it by the maximum value of that attribute. However, the disadvantage of this method is that if the dataset contains some attributes that has negative values some normalized values become negative. To overcome this problem we used a specific formula that ensures all the normalized values are between zero and one [21].

*C. Step 3: Creating Datasets from Each Level of Domain Ontology for Clustering*

In this step we create datasets for each level of the ontology one by one for each domain ontology that we created in 3.1. To make this sub task clear, in below description we take *TRO* as an example. The domain ontology for *TR* dataset includes three levels in the inclusion hierarchy. The twenty-four concepts at the lowest level of TRO represent the twenty four attributes of the original dataset. We get less number of attributes at higher levels when we move up from level 1 to level 3 as we generalise the lower level attributes to form higher level attributes. As there are 24 concepts at level 1 of TRO, and we take the 24 attributes of the original dataset at level 1 dataset. As TRO has 7 concepts at level 2, we convert the dataset for level 1 of TRO into a dataset with 7 attributes. For that we calculate the average values for each group of level 1 attributes and take them as the values of the level 2 that corresponds to the level 2 concept in the domain ontology. Here we use the average values of attributes so that each of the child attribute to give an equal weight so that each child attribute would have an equal influence on the clustering process. TRO has two attributes at level 3, the highest level. Again, we get the average of each group of level 2 attribute values as the value of the level 3.

*D. Step 4: Clustering the Datasets*

Once we prepare the three datasets for the three levels of each ontology, we perform clustering over the 3 levels of all the 6 datasets using the clustering algorithm GenClust++ [11]. GenClust++ has been proposed as an enhanced clustering algorithm that generates better clustering results compared to most of the currently available clustering algorithms. GenClust++ is available in the data mining environment WEKA [22] and clustering all the 18 datasets was done in WEKA. As GenClust++ is a parameter free clustering algorithm we don't have to specify the number of clusters and it automatically determines the number of clusters for a dataset.

*E. Step 5: Evaluating Clustering Results*

Clustering results from each dataset is evaluated to see whether there is an improvement in clustering quality. For that we first find out the records in each cluster resulted from the lowest level, level 1 of the domain ontology. Then, the clustering results from the datasets at higher levels, level 2 and 3 are compared with the same records in clustering results from the level 1 dataset. In this work, evaluation of clustering results is done using a cluster evaluation criterion called sum of squared error (SEE) [1, 21]. SEE value of each cluster resulted from each dataset for different levels of the ontology are calculated. This calculation is done on datasets derived from all the six original datasets.

## IV. EXPERIMENTAL RESULTS AND DISCUSSION

To evaluate the performance of our ontology-based clustering approach we select six numerical datasets from the UCI Machine Learning Repository [23]. Table IV shows a summary of the selected datasets. It includes name of each dataset, number of records, and number of attributes in each dataset. Out of these datasets Glass Identification (GI) dataset has the minimum number of attributes and records. Turkiye Student Evaluation (TKS) dataset has the maximum number of attributes and records.

TABLE IV. A SUMMARY OF THE SELECTED DATASETS

| # | Name | # of records | # of attributes |
|---|---|---|---|
| 1 | Diabetic Retinopathy Debrecen (DRD) | 1151 | 20 |
| 2 | Wine Quality (WQ) | 4898 | 12 |
| 3 | Glass Identification (GI) | 214 | 10 |
| 4 | Turkiye Student Evaluation (TKS) | 5820 | 33 |
| 5 | Image Segmentation (IS) | 2310 | 19 |
| 6 | Travel Review (TR) | 5456 | 25 |

### A. Experimental Results

We conduct evaluation on the clustering results from all the six datasets to evaluate our ontology-based clustering approach. We see whether, the clustering quality is improving or not from the lowest level to the highest level of a domain ontology. The evaluation is done independently on each of the 6 datasets. The clustering results we obtained on these six datasets are evaluated using SSE. According to this technique a cluster is considered to be good if the SSE value is lower. The SSE values we receive for the 3 levels of the 6 datasets, DRD, WQ, GI, TKS, IS and TR are shown in each column of Table V. Each row of this table includes SEE values for a specific level of each of the 6 domain ontologies.

TABLE V. IMPROVEMENT OF SSE (LOWER THE BETTER) IN SIX DATASETS.

| L | DRD | WQ | GI | TSK | IS | TR |
|---|---|---|---|---|---|---|
| L1 | 1775.73 | 3646.70 | 217.26 | 123524.09 | 4430.54 | **43877.41** |
| L2 | 1342.16 | 3287.27 | 217.26 | 132254.34 | 4424.24 | 77332.15 |
| L3 | **830.83** | **2451.24** | **176.36** | **121357.81** | **4419.81** | *73989.66* |

Then, we calculate the percentage of improvement in clustering quality based on the change in SEE value from lowest level to highest level. They are given in Table VI as two consecutive improvements and the total improvement in SSE value. That is the improvement from level 1 to level 2 and then from level 2 to level 3. The total percentage of improvement is an aggregate of the previous two improvements.

TABLE VI. IMPROVEMENT OF SSE (LOWER THE BETTER) AS A PERCENTAGE IN SIX DATASETS.

| Enhance | DRD | WQ | GI | TSK | IS | TR |
|---|---|---|---|---|---|---|
| L1 to L2 | 24.42 | 9.86 | 0 | -7.07 | 0.14 | -76.25 |
| L2 to L3 | 38.10 | 25.43 | 18.83 | 8.24 | 0.10 | 4.32 |
| Total | 62.51 | 35.29 | 18.83 | 1.17 | 0.24 | -71.92 |

In Table VI DRD dataset shows the highest percentage of improvement (62.51) in SSE value. WQ dataset has shown the second highest improvement of 35.29% while GI dataset has shown the 3$^{rd}$ highest improvement of 18.83%. The datasets TSK and IS have shown little improvements, 1.17 and 0.24. TSK dataset has shown a small overall improvement and an improvement from level 2 to 3, but a decline of clustering quality from level 1 to 2. Even though the last dataset TR have not shown an overall improvement in SEE it has shown a 4.32% improvement from level 2 to 3. All the datasets have shown some improvement in clustering quality at least at two levels. We observe a comparatively higher negative value of SEE in TR dataset from L1 to L2. The reason for this deviation is the higher number of clusters, 71 at level 1 resulted by GenClust++. The results of the clustering experiments for the 3 levels of the 6 datasets we conducted using GenClust++ are given in Table VII. Even though it is possible for us to get different number of clusters for different datasets for most of the datasets GenClust++ creates 2 to 4 clusters at each level.

TABLE VII. NUMBER OF CLUSTERS GENERATED BY GENCLUST++ FOR EACH LEVEL OF ONTOLOGY

| Level of Ontology | DRD | WQ | GI | TSK | IS | TR |
|---|---|---|---|---|---|---|
| L1 | 4 | 2 | 2 | 3 | 2 | 71 |
| L2 | 4 | 2 | 2 | 2 | 2 | 2 |
| L3 | 30 | 40 | 2 | 3 | 2 | 3 |

### B. Discussion

In our work, we investigate how clustering quality changes according to the level of the ontology. For that we create domain specific ontology for 6 numerical datasets that we obtained from the UCI Machine Learning Repository. According to the clustering experiments and the evaluation we conducted we can see a considerable improvement in cluster quality from lowest level to the highest level of domain ontology.

Many current ontology-based clustering techniques including [7, 12] have used WordNet as a reference ontology that is general in nature. WordNet ontology also includes many levels in its inclusion hierarchies. Domain ontology used in their work includes concepts like *Material Thing, Special Concept* and *Intangible* that are too general. Inclusion of general concepts would lead to put many or all the records in a single cluster that would hinder identifying clusters with domain specific characteristics. Due to this fact, in some work evaluation results have not been as they expected in the aggregations of low dimensional space of concepts [5].

In our work, we propose creating domain specific ontology from the attributes of a numerical dataset. As these attributes specially describe values in a dataset they can be considered as domain specific concepts. That helps us to avoid non-domain specific general concepts been counted in clustering. This alleviates the afore-mentioned problem of poor cluster quality due to general concepts. Dimensionality of attributes for clustering is reduced by ignoring too frequent and too infrequent concept occurrences [5]. However, these attributes could be more significant for clustering. In our method, we get the average of a group of lower level (child) concepts and introduce higher level (parent) concepts by moving up in the inclusion hierarchies of the ontology. These higher level concepts we use are domain specific and representative concepts of the domain. In their approach [5] they use a top-down approach to select concepts from an ontology. The algorithm that is used in this work [5] ignores some concepts that may be useful to improve clustering quality. However, in our approach we use a bottom-up approach.

## V. CONCLUSION

In this paper, we presented a domain ontology-based clustering approach. We aim at improving the cluster quality of a numerical dataset by incorporating domain ontology in clustering. We produced a domain ontology from the attributes of a numerical dataset. In our approach, for each ontology we had three levels, therefore we produced three datasets from a single ontology. We then clustered the datasets using a parameter free genetic algorithm-based clustering technique called GenClust++. The clusters are evaluated using SSE. The SSE values of our approach indicate that clustering quality gradually improves from lower to the higher levels of an ontology for most of the datasets. The first five out of the six datasets have shown 62.51%, 35.29%, 18.83%, 1.17% and 0.24% improvement of cluster quality respectively.

We argue that the cluster quality can further be improved with better understanding of the domain and creating highly domain specific ontology. This can be supported by the involvement of domain experts and a thorough study of the domain that would have a positive influence to the clustering results. In the future we expect to extend our work to perform further evaluation. Moreover, we expect to evaluate our approach with datasets that has higher number of attributes that would form domain ontologies with more than 3 levels of inclusion hierarchies. Again, we plan to evaluate cluster quality of our approach with different clustering algorithms and different evaluation criteria.